\documentclass[12pt,letter]{article}
\pdfoutput=1
\usepackage{graphicx, epsfig, color,cite}
\usepackage{amsmath}
\usepackage{amssymb}
\usepackage{float}
\usepackage{subfig}
\usepackage{hyperref}

\def\eslt{\not\!\!\!{E_T}}
\def\to{\rightarrow}

\def\bi{\begin{itemize}}
\def\ei{\end{itemize}}

\def\tchi{\tilde\chi}

\def\tf{\tilde f}

\def\tst{\tilde t}

\def\tg{\tilde g}

\def\alt{\lesssim}
\def\agt{\gtrsim}
\def\be{\begin{equation}}  
\def\ee{\end{equation}}  
\def\bea{\begin{eqnarray}}  
\def\eea{\end{eqnarray}}

\begin{document}
\begin{titlepage}
\begin{flushright}
OU-HEP-240730
\end{flushright}

\vspace{0.5cm}
\begin{center}
  {\Large \bf Decoding the gaugino code, naturally, at high-lumi LHC}\\
\vspace{1.2cm} \renewcommand{\thefootnote}{\fnsymbol{footnote}}
{\large Howard Baer$^{1}$\footnote[1]{Email: baer@ou.edu },
Vernon Barger$^2$\footnote[2]{Email: barger@pheno.wisc.edu} and
Kairui Zhang$^3$\footnote[5]{Email: kzhang89@wisc.edu}
}\\ 
\vspace{1.2cm} \renewcommand{\thefootnote}{\arabic{footnote}}
{\it 
$^1$Homer L. Dodge Department of Physics and Astronomy,
University of Oklahoma, Norman, OK 73019, USA \\[3pt]
}
{\it 
$^2$Department of Physics,
University of Wisconsin, Madison, WI 53706 USA \\[3pt]
}
{\it 
$^3$Department of Physics,
University of Nebraska, Lincoln, NE 68588 USA \\[3pt]
}

\end{center}

\vspace{0.5cm}
\begin{abstract}
\noindent

Natural supersymmetry with light higgsinos is most favored to emerge from
the string landscape
since the volume of scan parameter space shrinks to tiny volumes for
electroweak unnatural models.
Rather general arguments favor a landscape selection of soft SUSY breaking
terms tilted to large values, but tempered by the atomic principle: that
the derived value of the weak scale in each pocket universe
lie not too far from its measured value in our universe.
But that leaves (at least) three different paradigms
for gaugino masses in natural SUSY models: unified (as in nonuniversal
Higgs models), anomaly-mediation form (as in natural AMSB) and mirage
mediation form (with comparable moduli- and anomaly-mediated contributions). 
We perform landscape scans for each of these, and show they populate different, but overlapping, positions in $m(\ell\bar{\ell})$ and $m(wino)$ space.
The first of these may be directly measurable at high-lumi LHC
via the soft opposite-sign dilepton
plus jets plus $\eslt$ signature arising from higgsino pair production while
the second of these could be extracted from direct wino pair production leading
to same-sign diboson production.

\end{abstract}
\end{titlepage}

\section{Introduction}
\label{sec:intro}

Superstring compactification on a Calabi-Yau manifold leaves some remnant
$N=1$ supersymmetry (SUSY) in the resulting $4-d$ theory\cite{Candelas:1985en},
and the question then is: at which scale $Q<m_P$ (where $m_P$ is the reduced Planck mass) is SUSY broken at?
Phenomenologically, does SUSY play a role\cite{Witten:1981nf,Kaul:1981wp} in stabilizing the
measured value of the weak scale $m_{weak}\simeq m_{W,Z,h}\sim 100$ GeV
against blow-up to much higher mass scales due to quantum corrections to
$m_h$? Recent LHC search results\cite{Canepa:2019hph,ATLAS:2024lda} which require gluino masses
$m_{\tg}\agt 2.2$ TeV and top-squark masses $m_{\tst_1}\agt 1.1$ TeV (within the context of simplified models) suggest SUSY as a somewhat
disfavored mechanism for extending the Standard Model\cite{ATLAS:2024lda},
although this sentiment is based on theoretical
prejudices arising from pre-21st century physics.

Expectations for weak scale SUSY (WSS) changed in the 21st century with the
advent of the string landscape\cite{Bousso:2000xa,Susskind:2003kw}.
In the landscape picture, it was found that there exists an enormous number
(the number $10^{500}$ is an oft-quoted value\cite{Ashok:2003gk}
but much higher numbers are also entertained) of string flux compactification possibilities\cite{Douglas:2006es},
each leading to different $4-d$ laws of physics
(but still based on CY compactifications with their remnant $N=1$ SUSY in the
$4-d$ theory). Each of these vacua possibilities can be realized in the
eternally inflating multiverse leading to Weinberg's anthropic solution to the
cosmological constant problem\cite{Weinberg:1987dv}: if $\Lambda_{CC}$ is much larger than its
measured value, then each pocket universe would expand so quickly that
large-scale structure ({\it e.g.} galaxy condensation, and hence star formation)
would not occur, and hence the needed complexity for observers would not ensue.
Weinberg's scheme relies on a landscape comprised of SM-like $d=4$ models
where only $\Lambda_{CC}$ scans in the multiverse.

Arkani-Hamed {\it et al.} present arguments for a so-called friendly
landscape, wherein only superrenormalizable operators scan\cite{Arkani-Hamed:2005zuc}, and hence only
dimensionful quantities such as $\Lambda_{CC}$ and $m_{weak}$ scan on the
landscape. In this case, gauge groups, gauge couplings and Yukawa couplings
are instead determined by dynamics instead of by anthropics, yielding a
so-called predictive landscape. For models including WSS\cite{Baer:2006rs},
where the weak scale arises as a consequence of soft SUSY breaking,
we expect instead the overall SUSY breaking scale to scan in the
multiverse.
Since in string theory no scale is preferred over any other, the $F$-term
fields should be distributed as complex numbers whilst $D$-term breaking fields are distributed as real numbers over the decades of possibilities; this leads
to an expected power-law draw to large soft terms\cite{Douglas:2004qg,Susskind:2004uv,Arkani-Hamed:2005zuc}
\be
f_{SUSY}\sim m_{soft}^{2n_F+n_D-1}
\ee
within the string landscape. For the textbook case of SUSY breaking via a single
$F$-term, there is already a linear draw to large soft terms.

The draw to large soft terms must be tempered by the anthropic
requirement that electroweak symmetry breaking (EWSB) properly occurs,
and with no charge or color breaking (CCB) minima\cite{Baer:2016lpj}.
Also, for complex nuclei to be formed,
the magnitude of the pocket universe (PU) weak scale must lie within
the so-called ABDS window of values\cite{Agrawal:1997gf}
\be
m_{weak}^{OU}/2\alt m_{weak}^{PU}\alt (3-5)m_{weak}^{OU}
\ee
where $m_{weak}^{OU}$ is the weak scale as measured in our universe, while
$m_{weak}^{PU}$ is the derived weak scale in each pocket universe within
the greater multiverse:
\be
(m_Z^{PU})^2/2=\frac{m_{H_d}^2+\Sigma_d^d-(m_{H_u}^2+\Sigma_u^u)\tan^2\beta}{\tan^2\beta -1}-\mu^2\simeq-m_{H_u}^2-\mu^2-\Sigma_u^u(\tst_{1,2}) .
\label{eq:mzsPU}
\ee
The ABDS window provides a (statistical) upper bound on the soft SUSY
breaking terms.
By combining the draw to large soft terms with the ABDS window, and rescaling
$m_Z$ to its measured value in our universe, we can gain probability
distributions for Higgs boson and sparticle masses from the
landscape\cite{Baer:2017uvn}.
From a variety of well-motivated SUSY models (gravity-mediated\cite{Baer:2017uvn},
anomaly-mediated\cite{Baer:2023ech} and mirage-mediated\cite{Baer:2019tee} SUSY breaking), one finds
probability distributions which peak around $m_h\sim 125$ GeV with sparticles
somewhat or well-beyond LHC search limits. From this point of view,
LHC experiments have only now begun to explore the expected SUSY model
parameter space, and hence it is no surprise that SUSY has yet to be
revealed at LHC\cite{Baer:2020kwz}.

\section{Three paradigms for gaugino masses}
\label{sec:models}

The different dependences of the various soft SUSY breaking terms on the hidden
sector fields means that different soft terms should scan to large values
independently of each other\cite{Baer:2020vad}.
This leads to three broad expectations for sparticle mass spectra
oriented according to the relative values of the different gaugino
masses, referred to as the {\it gaugino code} in Ref.~\cite{Choi:2007ka}.

\subsection{Gravity-mediated SUSY breaking}

Under gravity mediation, scalar masses can be computed from the
supergravity K\"ahler potential $K(h_m^\dagger ,h_m)$, where the $h_m$
label hidden sector fields where SUSY breaking takes place.
In realistic compactifications, $K$ can be a complicated function of the
$h_m$, leading to {\it non-universal} scalar masses\cite{Soni:1983rm,Kaplunovsky:1993rd,Brignole:1993dj},
which may be regarded as a {\it prediction} from the sort of SUGRA theories
arising from string flux compactification.\footnote{The original mSUGRA model\cite{Arnowitt:1993qp} thought to arise
  from SUGRA models has an enforced ad-hoc implementation of scalar mass
  universality which violates this expectation.
  Thus, it is probably better to call that model constrained MSSM, or
  CMSSM\cite{Kane:1993td}, in that the universality condition goes against
  the expectations from supergravity.}
The ad-hoc scalar mass universality of mSUGRA/CMSSM models was implemented
in earlier days of SUSY phenomenology to enforce the super-GIM mechanism
to suppress flavor violation in SUSY models. Nowadays, universality forces
the CMSSM into EW unnaturalness, whereas non-universal Higgs masses
allow for radiatively-driven naturalness\cite{Baer:2012up,Baer:2012cf},
wherein large high scales values of $m_{H_u}^2$ are driven to natural
weak scale values.
In addition, the pull of the string landscape on first/second generation
scalar masses is to the 10-40 TeV level, but to flavor independent upper
bounds that arise from 2-loop RGE terms\cite{Arkani-Hamed:1997opn}.
This yields a landscape decoupling/quasi-degeneracy solution to the
SUSY flavor and CP problems\cite{Baer:2019zfl}.
Figuratively, scalar masses arise from operators such as
\be
\int d^2\theta d^2\bar{\theta} X^\dagger X Q_i^\dagger Q_j/m_P^2
\ee
where the hidden sector $X$ fields develop a SUSY breaking $F$-term
$F_X$, yielding scalar masses
$m_\phi^2\sim F_X^\dagger F_X /m_P^2$
where the gravitino mass $m_{3/2}^2\sim F_X^\dagger F_X/m_P^2\sim 1$ TeV for
$F_X\sim (10^{11}\ {\rm GeV})^2$. Thus, we expect non-universal
scalar masses of order $m_{3/2}$ in SUGRA theories\cite{Cremmer:1982en}.
But since the scalar fields of each generation fill out a complete
16-dimensional spinor rep of $SO(10)$, we expect universality of scalar
masses within each generation such that $m_0(1,2,3)$ are all independent
(this follows from local grand unification\cite{Buchmuller:2005sh} in string compactifications
and follows one of Nilles' golden rules of string phenomenology\cite{Nilles:2014owa}).
Thus, as far as soft SUSY breaking scalar masses go, we expect most plausibly
from string flux compactifications that
\be
m_0^2(1)\ne m_0^2(2)\ne m_0^2(3)\ne m_{H_u}^2\ne m_{H_d}^2
\ee
but with $m_0(1)\simeq m_0(2)\gg m_0(3)$.

The gaugino masses in SUGRA arise from the holomorphic
gauge kinetic function (GKF) $f_{AB}(h_m)$ where $A$ and $B$ are gauge indices.
For a simple GKF $f_{AB}\sim h_m\delta_{AB}$ then one gains universal
gaugino masses {\it e.g.}
\be
\int d^2\theta\frac{X}{m_P}W_AW^A\to \frac{F_X}{m_P}\lambda\lambda
\ee
with gaugino mass $m_\lambda\sim m_{3/2}$.

\subsection{Anomaly mediated SUSY breaking (AMSB)}

Notice in this case that since $f_{AB}$ is holomorphic that if hidden sector
fields leading to SUSY breaking have a surviving $R$-symmetry or other conserved quantum number,
then the gaugino masses will be forbidden at tree level, unlike scalar masses,
and then their leading contribution will be from the loop suppressed
anomaly-mediated terms\cite{Randall:1998uk,Giudice:1998xp,Bagger:1999rd}, where
\be
m_\lambda =\frac{\beta (g)}{g}m_{3/2}
\ee
where $\beta$ is the beta-function of the associated gauge group.
Then we expect
\bea
M_1&=& \frac{33}{5}\frac{g_1^2}{16\pi^2}m_{3/2}\\
M_2&=& \frac{g_2^2}{16\pi^2}m_{3/2}\ \ \ {\rm and} \\
M_3&=& -3\frac{g_3^2}{16\pi^2}m_{3/2}
\eea
where the loop suppression from $m_{3/2}$ is evident.
Thus, in such models one can expect heavier scalar masses of order
$\sim m_{3/2}\sim 10$ TeV whilst gaugino masses may be at or below
the TeV scale, but of AMSB form. This scenario is
termed PeV-scale SUSY by Wells\cite{Wells:2004di} and
split\cite{Arkani-Hamed:2004zhs} and minisplit\cite{Arvanitaki:2012ps}
(depending on the magnitude of scalar masses) and can be parametrized
within the {\it minimal} AMSB model with
parameters\cite{Gherghetta:1999sw,Feng:1999hg}:
\be
m_0,\ m_{3/2},\ \tan\beta,\ sign (\mu )\ \ \ \ (mAMSB)
\ee
where $m_0$ is an ad-hoc {\it bulk} scalar mass introduced to
avoid the problem of tachyonic slepton masses in the case of pure
anomaly-mediation\cite{Randall:1998uk}.
Famously, mAMSB models feature a wino as lightest SUSY particle (LSP).
A wino LSP seems ruled out by direct and indirect WIMP search
experiments\cite{Cohen:2013ama,Fan:2013faa,Baer:2016ucr}.
Also, requiring $m_h\sim 123-127$ GeV, in Ref. \cite{Baer:2014ica}
it was found that mAMSB was highly EW finetuned over all remaining 
parameter space. By extending the bulk scalar mass contributions to
include independent Higgs terms $m_{H_u}^2$ and $m_{H_d}^2$, and by including
bulk $A$ terms $A_0$, then the AMSB model could be generalized\cite{Baer:2018hwa} such
that $m_h\sim 125$ GeV was allowed along with EW naturalness,
although in the EW-natural regions, the higgsinos were the lightest
EWinos, while the wino was still the lightest of the gauginos.
This natural generalized AMSB model\cite{Baer:2018hwa} is thus parametrized by
\be
m_0(i),\ m_{3/2},\ A_0,\ \tan\beta, \mu ,\ m_A\ \ \ (nAMSB) 
\ee
where $m_{H_u}^2$ and $m_{H_d}^2$ have been traded for the more convenient
weak scale parameters $\mu$ and $m_A$.

\subsection{Comparable gravity and anomaly mediation (mirage mediation)}

The mirage mediated form of soft SUSY breaking terms was originally
derived\cite{Choi:2005ge} within the context of KKLT\cite{Kachru:2003aw}
flux compactifications of type IIB string theory on a Calabi-Yau
orientifold.
In KKLT, background gauge fields in the 6-d compact space are given quantized
flux values which thread various cycles, leading to stabilization of
the dilaton and all complex structure moduli fields $Z_\alpha$ and
gain masses of order the ultra-high Kaluza-Klein (KK) scale. 
The remaining K\"ahler moduli $T_\beta$ are assumed to be stabilized via
non-perturbative effects such as gaugino condensation\cite{Nilles:1990zd},
and gain much smaller masses.
At this stage the model has unbroken $N=1$ SUSY with an AdS vacuum.
To obtain a broken SUSY theory with deSitter vacuum, 
KKLT assumed a SUSY breaking anti-$D3$ brane at the tip of a
Klebanov-Strassler\cite{Klebanov:2000hb} throat which gave an uplift to a metastable dS vacuum.
In this setup, it was pointed out by Choi {\it et al.}\cite{Choi:2005ge}
that the construct allowed for weak scale SUSY as a solution to the GHP,
but with a little hierarchy
\be
m_T\gg m_{3/2}\gg m_{soft}
\label{eq:hierarchy}
\ee 
where the proportionality factor between terms in Eq. \ref{eq:hierarchy}
was a factor $\sim\log (m_P/m_{3/2})\sim 4\pi^2$. 
In such a case, the gravity/moduli-mediated soft terms
were $\sim m_{soft}$ while the anomaly-mediated contributions were
$m_{AMSB}\sim m_{3/2}/16\pi^2$, suppressed by a loop factor but now comparable
to the gravity-mediated contributions due to the hierarchy in
Eq. \ref{eq:hierarchy}.

The original derivation of soft terms is given in Ref. \cite{Choi:2005ge},
which combined gravity- and anomaly-mediated contributions in a scenario
which assumed a single K\"ahler modulus,
with visible fields on a D3 or D7 brane. 
This class of models is known as mirage mediation since the
gaugino masses start off at $Q=m_{GUT}$ with non-universal values,
but which evolve to a unified value at some intermediate
scale\cite{Falkowski:2005ck}
\be
\mu_{mir} =m_{GUT} e^{(-8\pi^2/\alpha )}
\ee
where the RG evolution offsets the displacement of gaugino masses by 
the AMSB term, which includes the gauge group beta function.
Here, $\alpha$ parametrizes the relative AMSB vs. gravity-mediated contributions to gaugino masses.
These models, with discrete values of gravity-mediated soft terms, were found 
to be highly unnatural under the conservative finetuning measure $\Delta_{EW}$
for $m_h\sim 125$ GeV\cite{Baer:2014ica}.

However, in Ref. \cite{Baer:2016hfa} it was proposed that the soft terms could be generalized to a form
which ought to hold under more realistic compactifications
where the K\"ahler moduli could number in the hundreds. This generalized
mirage mediation model (GMM) has soft terms given by
\bea
M_a &=& (\alpha +b_ag_a^2) m_{3/2}/16\pi^2 ,\ \ \ (a=1-3)\\
A_\tau &=& (-a_3\alpha +\gamma_{L_3}+\gamma_{H_d}+\gamma_{E_3}) m_{3/2}/16\pi^2,\\
A_b &=& (-a_3\alpha +\gamma_{Q_3}+\gamma_{H_d}+\gamma_{D_3}) m_{3/2}/16\pi^2,\\
A_t &=& (-a_3\alpha +\gamma_{Q_3}+\gamma_{H_u}+\gamma_{U_3}) m_{3/2}/16\pi^2,\\
m_i^2(1,2) &=& (c_{m}\alpha^2+4\alpha\xi_{i}-\dot{\gamma}_{i}) m_{3/2}/16\pi^2,\\
m_j^2(3) &=& (c_{m3}\alpha^2+4\alpha\xi_{j}-\dot{\gamma}_{j}) m_{3/2}/16\pi^2,\\
m_{H_u}^2 &=& (c_{H_u}\alpha^2+4\alpha\xi_{H_u}-\dot{\gamma}_{H_u}) m_{3/2}/16\pi^2,\\
m_{H_d}^2 &=& (c_{H_d}\alpha^2+4\alpha\xi_{H_d}-\dot{\gamma}_{H_d}) m_{3/2}/16\pi^2
\eea
where $\alpha$ parametrizes the relative modulus/AMSB mixing in the gaugino masses, $b_a$ are the gauge $\beta$-function coefficients for gauge group $a$ and the $g_a$ are the corresponding gauge couplings.
Also, the parameters $a_3\equiv a_{Q_3 H_u U_3}$ etc.,
$c_{m1}$, $c_{m2}$, $c_{m3}$, $c_{H_u}$ and $c_{H_d}$ are all now generalized
to be continuously variable.
Furthermore, $\xi_i =\sum_{j,k}a_{ijk}\frac{y_{ijk}^2}{4}-\sum_ag_a^2C_2^a(f_i)$
where the $y_{ijk}$ are the superpotential Yukawa couplings, the $C_2^a$ is the quadratic Casimir for the $a$th gauge group corresponding to the representation to which the sfermion $\tf_i$ belongs, $\gamma_i$ is the anomalous dimension
and $\dot{\gamma}_i=8\pi^2\frac{\partial\gamma_i}{\partial\log\mu}$.
Expressions for the last two quantities involving anomalous dimensions
can be found in the Appendices of Ref. \cite{Falkowski:2005ck}. 

The generalized MM parameter space is thus given by
\be
\alpha,\ m_{3/2},\ c_{mi},\ a_3,\ c_{H_u},\ c_{H_d},\ {\rm and}\ \tan\beta\ \ \ (GMM)
\ee
where $i=1-3$ runs over the generations. The independent values of $c_{H_u}$ and $c_{H_d}$, which set the modulus-mediated contribution to the Higgs soft masses,
can be traded for the more convenient weak scale values of $\mu$ and $m_A$.
Thus, the final parameter space is given by
\be
\alpha,\ m_{3/2},\ c_{mi},\ a_3,\ \mu,\ m_A\ {\rm and}\ \tan\beta\ \ \ (GMM^\prime)
\ee
where $i$ is a generation index.
For convenience, we will also take the first two generation $c_{m1}=c_{m2}$ since
these are pulled to large values $\sim 10-40$ TeV in the landscape
which leads to a mixed quasi-degeneracy/decoupling solution to the SUSY
flavor problem\cite{Baer:2019zfl}.

\subsection{The gaugino code in natural SUSY}

\begin{figure}[htb!]
\centering
    {\includegraphics[height=0.4\textheight]{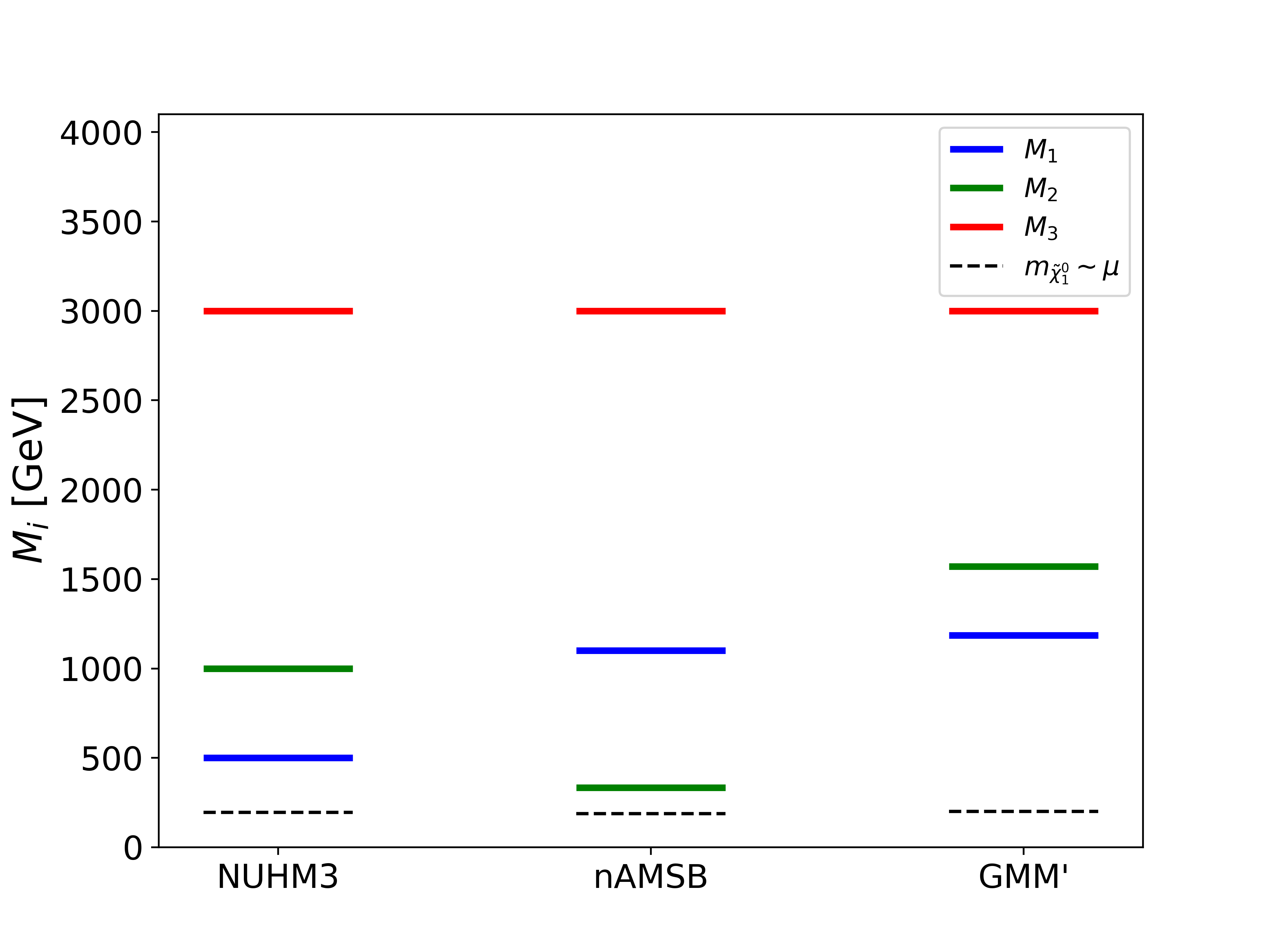}}
    \caption{Figurative plot of gaugino/higgsino masses in the
      three paradigm natural SUSY models with $m_{\tg}\simeq 3$ TeV
      and $\mu =200$ GeV.
      \label{fig:inomass}}
\end{figure}

The three possibilities for natural SUSY spectra are displayed in
Fig. \ref{fig:inomass}, which displays the gaugino code as
emphasized in Ref. \cite{Choi:2007ka}.
For all three cases, we expect in the string landscape
picture that first/second generation scalars will be drawn to the 10-40 TeV
range (thus solving the SUSY flavor/CP problems) whilst light stops
lie in the range $m_{\tst_1}\sim 1-3$ TeV and heavier stops
$m_{\tst_2}\sim 3-8$ TeV\cite{Baer:2017uvn,Baer:2019tee,Baer:2023ech}.
Also, we expect the (SUSY conserving) higgsino
mass terms to lie in the natural range $m_{\tchi_{1,2}^0,\tchi_1^\pm}\sim 100-350$
GeV. The distinguishing characteristic between the three models lies
with the orientation of the gaugino masses, as figuratively displayed
in Fig. \ref{fig:inomass}.
Supergravity models with a simple linear form of gauge kinetic function (GKF)
$f_{AB}\sim \delta_{AB}S$ or $\delta_{AB}T$ (where the singlet hidden sector
field $S$ is a dilaton and $T$ is a K\"ahler modulus) give
rise to unified gaugino masses at scale $Q\sim m_{GUT}$ while
RG evolution leads to the ratios $M_3:M_2:M_1\sim 6:2:1$ at
$Q=m_{weak}$ as is typical of the three extra parameter non-universal Higgs model
(NUHM3) of Fig. \ref{fig:inomass}. If the SUSY breaking fields are charged,
perhaps under an $R$-symmetry, then the linear form of the GKF is forbidden,
and the dominant contribution to gaugino masses comes from the loop-suppressed
AMSB form with $M_3:M_2:M_1\sim 9:1:3.3$ as shown in the nAMSB column.
Alternatively, if the gaugino masses obtain comparable gravity- and AMSB-
contributions as in mirage mediation models where $m_T\gg m_{3/2}\gg m_{soft}$,
then we expect a more compressed form of gaugino masses which unify at the
mirage scale $\mu_{mir}$ which would be intermediate between the weak and
GUT scales. This pattern is shown in the GMM' column.
While each of the three cases leads to its own orientation for gaugino masses,
we expect from naturalness that the lightest EWinos will be the
triplet of higgsino states $\tchi_{1,2}^0$, $\tchi_1^\pm$,
as denoted by the grey dashed lines.
The question then becomes: can experiments at (high-luminosity) LHC
distinguish between these three cases, thus decoding the gaugino code.

\section{Revealing the gaugino code, naturally, at high-lumi LHC}
\label{sec:decoding}

The LHC has two promising avenues towards decoding the natural
gaugino code, namely by observing 1. higgsino pair production and
2. wino pair production.
It turns out each of these channels is quite distinctive to the case of
natural SUSY models and signals should gradually emerge as LHC accrues
higher and higher integrated luminosity. From Run 2 data, both ATLAS
and CMS seem to have small $2\sigma$ excesses in the higgsino pair
production channel\cite{ATLAS:2019lng,CMS:2021edw}.
It is thus of great interest to see if these excesses
build up or fade away with the Run 3 and HL-LHC data sets.

\subsection{Higgsino pair production}

Charged higgsino pair production $pp\to \tchi_1^+\tchi_1^-$ followed by
$\tchi_1^\pm\to\tchi_1^0 f\bar{f}'$ decay (where $f$ stands for SM fermions)
typically yields very
soft visible decay products and $\eslt$ and a rather indistinct signature.
Neutral higgsino pair production\cite{Baer:2011ec}
\be
pp\to\tchi_2^0\tchi_1^0\ \  {\rm with}\ \ \ \tchi_2^0\to\tchi_1^0 f\bar{f}
\label{eq:osdjmet}
\ee
may also lead to very low visible energy,
unless the scattering reaction recoils from a hard initial state
quark/gluon radiation\cite{Han:2014kaa,Baer:2014kya,Han:2015lma} (ISR).
In this latter case,
letting $f$ be $e$ or $\mu$, then we expect soft opposite-sign dilepton plus
jet plus $\eslt$ production (OSDLJMET) where the opposite-sign (OS)
dilepton pair has invariant mass $m(\ell\bar{\ell})$ kinematically
bounded by $m_{\tchi_2^0}-m_{\tchi_1^0}$. Thus, in the signal $m(\ell\bar{\ell})$
distribution, we expect an excess beyond SM expectations for
$m(\ell\bar{\ell})< m_{\tchi_2^0}-m_{\tchi_1^0}$ while data should agree
with SM expectations for higher invariant dilepton masses.\cite{Kitano:2006gv}
Along with offering a distinctive signal channel for natural SUSY models
at HL-LHC, this channel also provides a direct measure of the neutral
higgsino mass difference $m_{\tchi_2^0}-m_{\tchi_1^0}$.
A rough estimate of the superpotential $\mu$ parameter may also be gleaned
from this channel based on total rate since
$m_{\tchi_2^0}\sim m_{\tchi_1^0}\sim \mu$ and the total production rate
mainly depends on the raw higgsino masses. In fact, the exclusion limits
are typically plotted in the $m_{\tchi_2^0}$ vs $m_{\tchi_2^0}-m_{\tchi_1^0}$
plane\cite{Baer:2020sgm}.
More refined cuts for extracting the OSDLJMET signal have recently been
presented in Ref's. \cite{Baer:2021srt,Baer:2022qrw}.

\subsection{Wino pair production}

Another distinctive SUSY discovery channel which is endemic to natural
SUSY models is the clean same-sign diboson signal (SSdB) which arises from
wino pair production:
\be
pp\to\tchi_2^+\tchi_{3,4}^0\ \ {\rm with}\ \ \tchi_2^+\to W^+\tchi_{1,2}\ \ {\rm and}\ \ \tchi_{3,4}\to W^\pm\tchi_1^\mp
\ee
where the $\tchi_2^\pm$ is the charged wino and $\tchi_3^0$ is the neutral wino
in nAMSB or $\tchi_4^0$ is the neutral wino in GMM' or NUHM3.
Since the visible decay products of the daughter higgsinos are very soft,
and upon leptonic decays of the final state $W$ bosons, this reaction leads
half the time to a hadronically clean (aside from usual ISR) same-sign
dilepton $+\eslt$ final state which has very low SM
backgrounds\cite{Baer:2013yha,Baer:2013xua,Baer:2017gzf}.
This is different from the older same-sign dilepton signals arising from
gluino and squark pair production which would be accompanied by
multiple hard jets\cite{Baer:1988qp,Barnett:1988mx,Baer:1989hr,Barnett:1993ea}.

Due to the required two $W$-boson leptonic branching fractions,
the rate for this channel for the current LHC Run 2 data set is
typically expected around the 1-event level,
so at present no search limits are available from this channel.
However, as more integrated luminosity accrues, this channel will become
more and more important due to the extremely low SM background levels.
Since the production rate mainly depends on $m(wino)$,
and the wino branching fractions to $W$ bosons is well-known
and theoretically stable, the direct production rate can yield a
measure of the gaugino mass $M_2$\cite{Baer:2017gzf}.
In Ref. \cite{Baer:2017gzf}, it is estimated that with
3 ab$^{-1}$ of integrated luminosity (corresponding to
$\sim 10-100$ events), HL-LHC should be able to determine
$m(wino)$ to around the  5-10\% level based on signal rate alone,
including statistical errors.

\subsection{Decoding the gaugino code at HL-LHC}

Next, we perform string landscape scans of soft SUSY breaking parameters
assuming the textbook case of an $n=1$ power-law landscape draw to
large soft terms.
Each class of soft terms (gaugino mass, scalar mass, $A$-terms) should
scan independently on the landscape owing to the different functional
dependence of soft terms on hidden sector fields for different string
compactification possibilities\cite{Baer:2020vad}.
Thus, we scan NUHM3 over the range\footnote{We use Isajet 7.91\cite{Paige:2003mg} for
sparticle and Higgs mass and $\Delta_{EW}$ generation.}
\bi
\item $m_0(1, 2): 0.1-60$ TeV,
\item $m_0(3): 0.1-20$ TeV,
\item $m_{1/2}: 0.5-10$ TeV,
\item $A_0: -50-0$ TeV (negative only),
\ei
at first for fixed $\mu =200$ GeV (since $\mu$ is not a soft term but arises
from whatever solution to the SUSY $\mu$ problem is assumed).\footnote{Twenty solutions to the SUSY $\mu$ problem are reviewed in Ref. \cite{Bae:2019dgg}.}
For each parameter set, the expected value of the weak scale in each pocket
universe is computed from $m_{weak}^{PU}=m_Z\sqrt{\Delta_{EW}/2}$ and
solutions with $m_{weak}^{PU}> 4m_{weak}^{OU}$ are rejected as lying beyond the
ABDS window. (An important technical point is that the scan range upper limits must be selected to be beyond the upper limits imposed by the anthropic
ABDS window.) To gain probability distributions for Higgs and sparticle masses
in our universe, then as a final step $m_Z^{PU}$ must be adjusted
(but not finetuned!) to our value $m_Z^{OU}=91.2$ GeV.

For the nAMSB model, our landscape scan is over\cite{Baer:2023ech}
\bi
\item $m_{3/2}:\ 80-400$ TeV,
\item $m_0(1, 2):\ 1-20$ TeV,
\item $m_0(3):\ 1-10$ TeV,
\item $A_0:\ 0- 20$ TeV (positive only).
\ei

For GMM', our scan follows Ref. \cite{Baer:2019tee} with fixed $m_{3/2} = 20$ TeV:
\bi
\item $\alpha:\ 3-25$ (with $\alpha^1$ draw),
\item $c_{m1,2}= (16\pi^2/\alpha)^2$
\item $\sqrt{c_{m3} \alpha^2}:\ 3-80$ and
\item $(a_3 \alpha):\ 3-100$
\ei
For all cases, we scan with a linear draw $m_A:0.25-10$ TeV and
uniformly over $\tan\beta:\ 3-60$.

\begin{figure}[htb!]
\centering
    {\includegraphics[height=0.4\textheight]{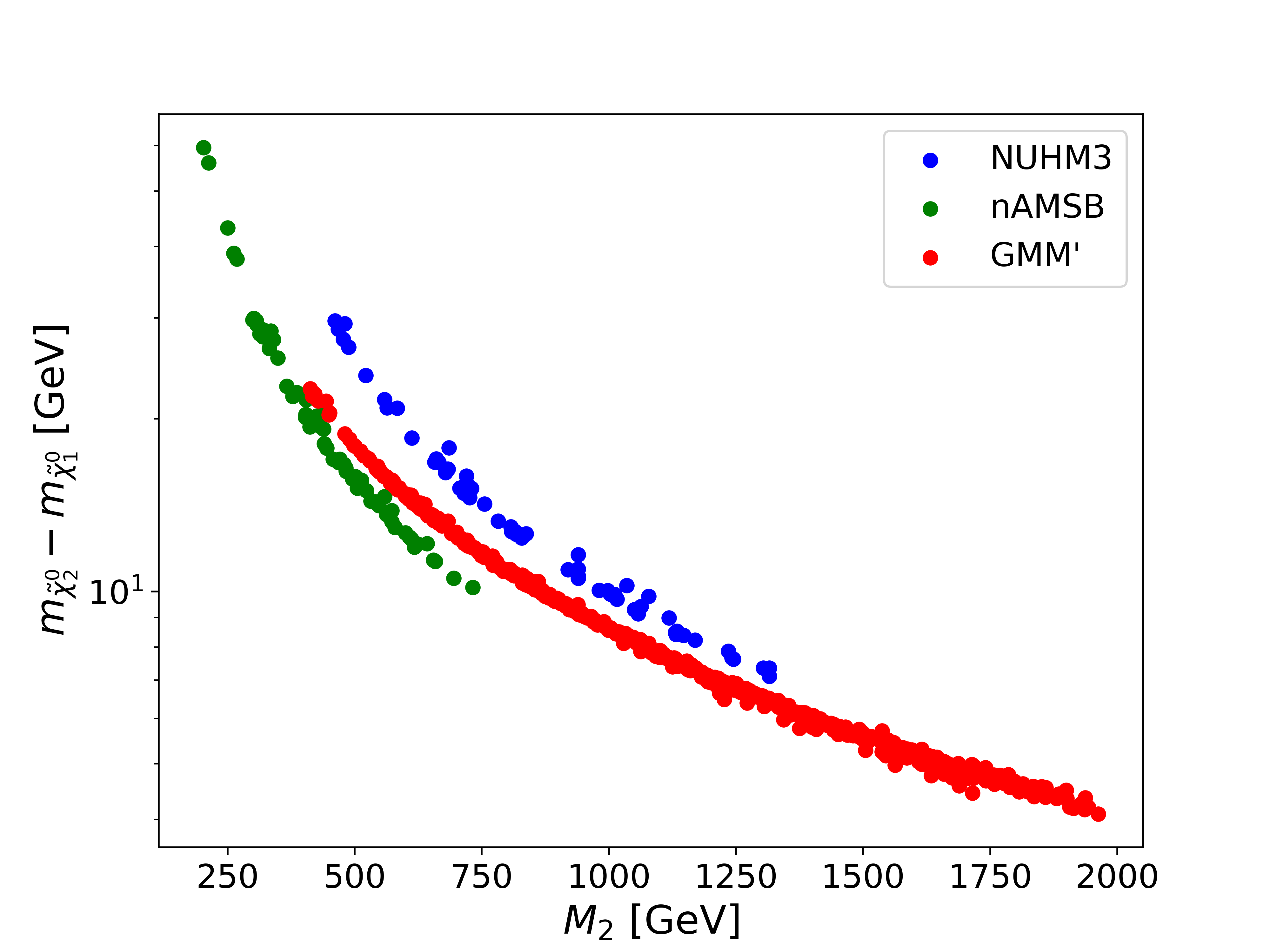}}
    \caption{Locus of landscape scan points from the NUHM3 model, the
      nAMSB model and GMM' model in the $m_{\tchi_2^0}-m_{\tchi_1^0}$
      vs. $m_{\tchi_2^+}\simeq M_2$ plane for fixed value $\mu =200$ GeV.
      \label{fig:plane1}}
\end{figure}

The results of our scans are plotted in Fig. \ref{fig:plane1} in the
$\Delta m^0\equiv m_{\tchi_2^0}-m_{\tchi_1^0}$ vs. $M_2$ plane.
Green dots denote nAMSB points, blue dots denote NUHM3 points and
red dots denote GMM' points. From Fig. \ref{fig:plane1}, we see the
three models inhabit somewhat different locusses in the plane.
The green nAMSB points extend to rather large $\Delta m_0$ values
--as high as $\Delta m^0\sim 60$ GeV-- since in this case the light winos
can mix with the higgsinos, thus breaking the expected inter-higgsino mass
degeneracy. Also, the range in $M_2$ for nAMSB points only extends as high
as $\sim 750$ GeV. In contrast, the NUHM3 model with unified gaugino
masses has a maximal $\Delta m^0\sim 30$ GeV but $M_2$ can extend
as high as $\sim 1250$ GeV where $\Delta m^0\sim 7$ GeV.
The GMM' is intermediate between these models, since it has mixed gravity/AMSB
mediation. At low $M_2$, where $\alpha$ is small, it looks more like AMSB
spectra and has a slight overlap with the green points, while at high
$\alpha$ values it looks more like gravity-mediation and lies closer to the
blue points, although the compressed GMM' gaugino spectrum allows
$M_2$ to extend much further, as high as $M_2\sim 2$ TeV where
$\Delta m^0$ is as low as $\sim 4$ GeV. Notice in all cases $\Delta m^0$
never reaches below the few GeV case, since for that to occur, $M_2$
values would have to be so high that $\Sigma_u^u(m(wino))$ would become
unnatural\cite{Baer:2015rja}, and lie outside the ABDS window.

\begin{figure}[htb!]
\centering
    {\includegraphics[height=0.4\textheight]{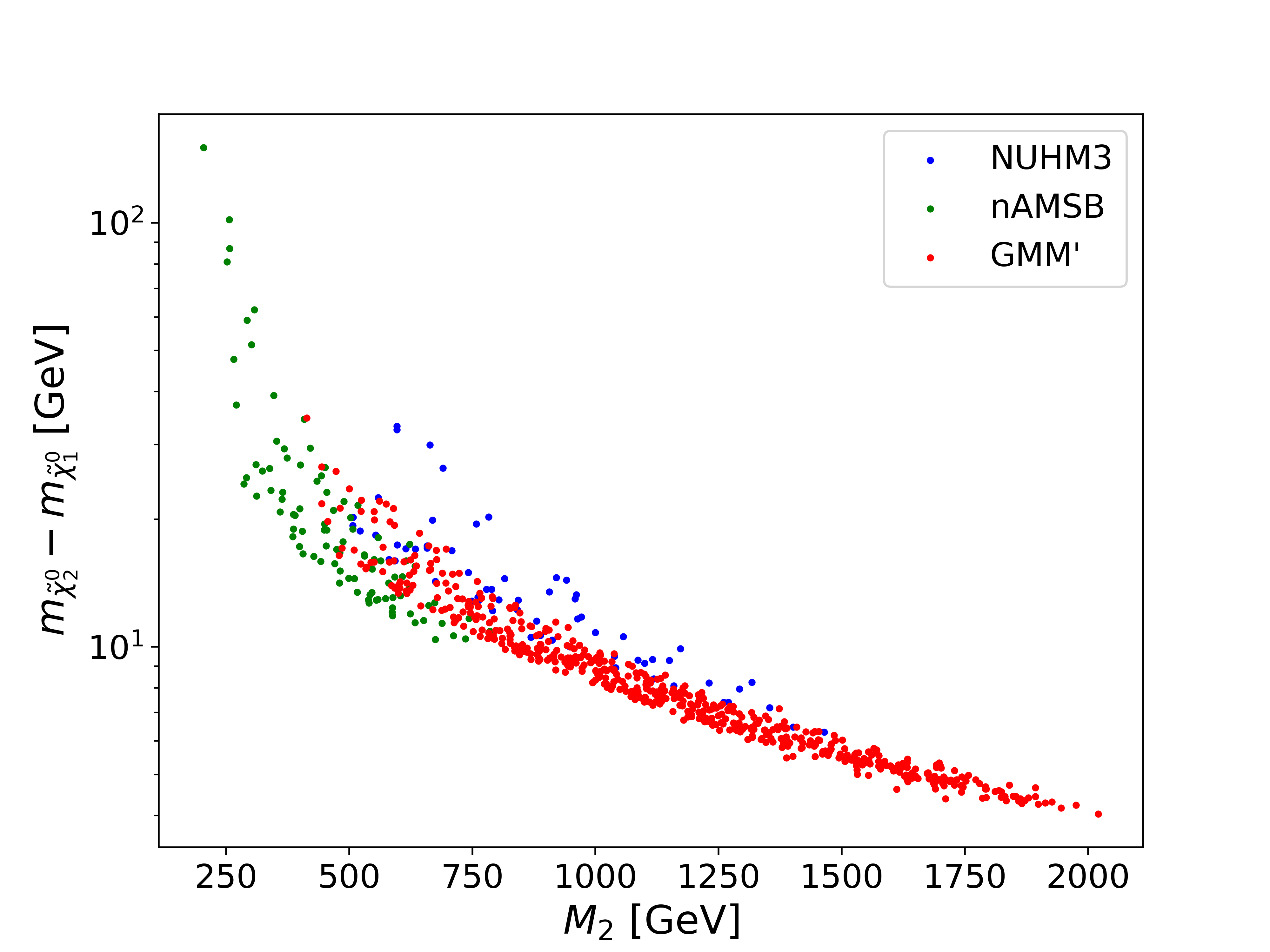}}
    \caption{Locus of landscape scan points from the NUHM3 model, the
      nAMSB model and GMM' model in the $m_{\tchi_2^0}-m_{\tchi_1^0}$
      vs. $m_{\tchi_2^+}\simeq M_2$ plane for a uniform scan over
      $\mu =100-350$ GeV.
      \label{fig:plane2}}
\end{figure}

In Fig. \ref{fig:plane2}, we again plot the locus of $n=1$ landscape
scan points in the $m_{\tchi_2^0}-m_{\tchi_1^0}$ vs. $M_2$ plane, but this time
including a uniform scan over $\mu :100-350$ GeV, in case the $\mu$ parameter
cannot be well-measured. In this case, the model scan points broaden out
and have some substantial overlap in the intermediate mass region.
Nonetheless, the landscape scan regions still maintain some degree of
separation as shown by the colored dots, so depending on where the
ultimate measure values of $\Delta m^0$ and $m_{\tchi_1^\pm}$ lay, the
locus may or may not lie within a distinguishable region. Clearly, a
more precide measured value of $\mu$ will help aid in this enterprise.

\section{Conclusions}
\label{sec:conclude}

String flux compactifications on a Calabi-Yau manifold with special holonomy
preserve some remnant $N=1$ supersymmetry which provides
a 't Hooft technical natural solution to the big hierarchy problem.
The question then is: what is the scale of SUSY breaking and does the subsequent
SUSY spectrum avoid the little hierarchy problem based on practical
naturalness? Rather general arguments from the string landscape favor a
power-law draw to large soft terms tempered by the anthropic requirement
that the derived value of the weak scale lie within the so-called ABDS window.
In this case, EW {\it natural} SUSY models are most likely to emerge as
opposed to finetuned SUSY models since the latter viable parameter space
shrinks to a relatvely tiny volume compared to natural models
(one must happen upon the finetuned parameters that lead to a weak scale
not-too-far removed from its measured value in our universe.
These natural models would be characterized by first/second generation scalars
in the 10-40 TeV range while top-squarks inhabit the few TeV range and
higgsinos lie not to far from $m_{weak}$ in the 100-350 GeV range.
Within the context of natural models, three main paradigms are viable,
distinguished by the orientation of the gaugino masses: unified
gaugino masses as in the NUHM3 model, AMSB gaugino masses as in nAMSB
and mirage mediated gaugino masses as in the GMM' model.

Can forthcoming data from HL-LHC distinguish these cases,
thus decoding the gaugino code? 
Two avenues to detection of natural SUSY models lie in the ODLJMET signal
from higgsino pair production and the SSdB signature arising from wino pair
production. The first admits a precise measurement of
$\Delta m^0=m_{\tchi_2^0}-m_{\tchi_1^0}$ via the expected edge in the
$m(\ell\bar{\ell})$ distribution while wino pair production may allow
extraction of the wino mass $M_2$ from the total rate of SSdB production.
By placing these measurements within the $\Delta m^0$ vs. $M_2$ plane,
one may well be able to decode the gaugino code since SUSY models
predict the different cases may inhabit distinctive regions of this
parameter plane

{\it Acknowledgements:} 

VB gratefully acknowledges support from the William F. Vilas estate.
HB gratefully acknowledges support from the Avenir Foundation.


\bibliography{inocode}
\bibliographystyle{elsarticle-num}

\end{document}